\documentclass{article}

\usepackage{PRIMEarxiv}

\usepackage[utf8]{inputenc} 
\usepackage[T1]{fontenc}    
\usepackage[hidelinks]{hyperref}       
\usepackage{url}            
\usepackage{booktabs}       
\usepackage{amsfonts}       
\usepackage{nicefrac}       
\usepackage{microtype}      
\usepackage{lipsum}
\usepackage{fancyhdr}       
\usepackage{graphicx}       
\graphicspath{{media/}}     

\usepackage{algorithm}
\usepackage{algpseudocode}
\usepackage{booktabs}
\usepackage{tabularx}
\usepackage{array}

\usepackage{amsmath, amssymb}

\usepackage{color}
\usepackage{soul}
\usepackage{xcolor}

\newcommand{\CaDiCaL}{\textsc{CaDiCaL}}

\title{Queen Domination by SAT Solving}

\author{
  Taha Rostami \\
  University of Luxembourg \\
  \texttt{taha.rostami@uni.lu}
  \And
  Curtis Bright \\
  University of Waterloo \\
  \texttt{cbright@uwaterloo.ca}
}

\begin{document}
\maketitle

\begin{abstract}
The queen domination problem asks for the minimum number of queens required to attack all squares on an $n \times n$ chessboard. Once this optimal number is known, determining the number of distinct solutions up to isomorphism has also attracted considerable attention. Previous work has introduced specialized and highly optimized search procedures to address open instances of the problem. While efficient in terms of runtime, confidence in their computational results ultimately depends on the correctness of these specialized implementations, motivating approaches that additionally provide independently verifiable correctness certificates. To this end, we present a proof-producing SAT framework for the queen domination problem based on an encoding that introduces auxiliary variables representing whether rows, columns, diagonals, and anti-diagonals contain queens, thereby exposing the geometric structure of the problem to the SAT solver. The framework is further strengthened through a novel literal-ordering strategy, symmetry breaking, a modern Cube-and-Conquer framework, and a unified proof-generation and verification pipeline. Together, these techniques yield both high performance and independently verifiable correctness. Our results uncover and correct a discrepancy in the previously reported enumeration for $n=16$ and resolve the previously open case $n=19$.
\end{abstract}


\section{Introduction}

\begin{figure}
    \centering
    \includegraphics[width=0.49\linewidth]{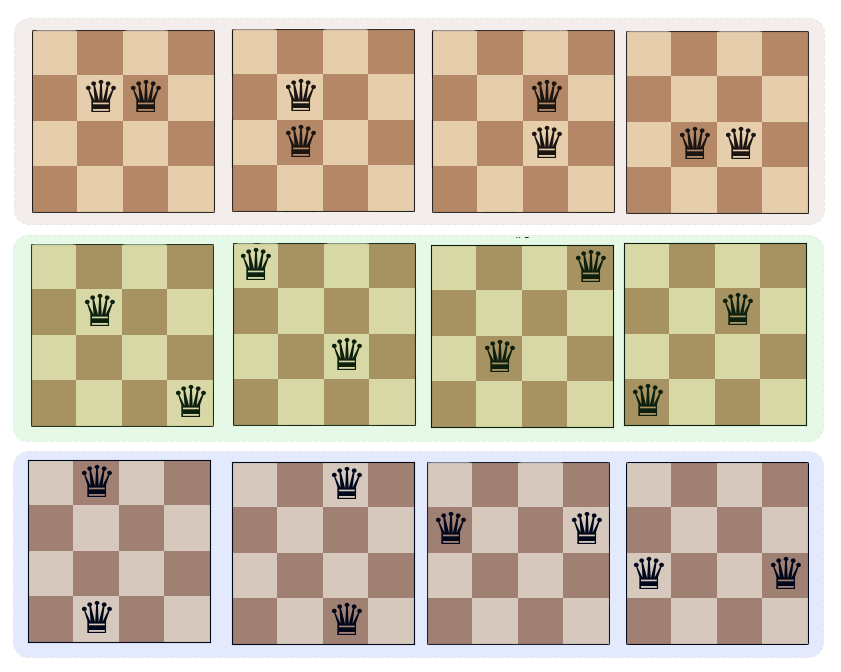}
    \caption{
A minimum of two queens is necessary to dominate all squares of a $4\times4$ chessboard. While there exist 12 distinct configurations that achieve this, many are symmetric under rotation and reflection. Up to isomorphism, these reduce to only three essentially different solutions, as highlighted in the figure.
    }
    \label{fig:n4-solutions}
\end{figure}

Given an $n \times n$ chessboard, the queen domination problem asks for the minimum number of queens, denoted $\gamma(Q_n)$, required to cover all squares of the board~\cite{hedetniemi2020}. A queen in chess is a piece that covers all squares in the same row, column, and diagonals from its position. For over a century and a half, mathematicians have studied not only the queen domination problem itself but also related variants~\cite{hedetniemi2020}. For example, in addition to determining the optimal value of $\gamma(Q_n)$, researchers have been interested in enumerating all distinct solutions of an $n\times n$ chessboard up to isomorphism~\cite{bird2017}. The latter is the primary focus of this paper. Figure~\ref{fig:n4-solutions} provides a concrete illustration of the minimum queen domination problem on a $4\times4$ chessboard. It presents all 12 optimal solutions and highlights their classification into three isomorphism classes based on board symmetries.

Before the advent and widespread use of computers in mathematics, early efforts to study the problem focused on establishing tighter theoretical bounds~\cite{cockayne1990, raghavan1987, grinstead1991} and solving small open cases manually~\cite{jaenisch1862}. Two theoretical observations are particularly useful for constraining the queen domination problem: the inequality $\gamma(Q_n)-\gamma(Q_{n-1})\leq1$ restricts how quickly the minimum number of queens can increase as $n$ grows, while the lower bound $\gamma(Q_n)\geq\lceil(n-1)/2\rceil$ further narrows the range of candidate values~\cite{cockayne1990}. Together, these bounds often reduce the number of values that need to be checked for a given $n$ to just one or two.

Theoretical work continues to this day~\cite{weakley2018}. Meanwhile, previously open cases have increasingly been tackled using algorithmic and computer-based search~\cite{bird2017, gibbons1997, kearse2000}. This has led to a fruitful synergy between theory and computation: theoretical results narrow the search space, while computational results provide new data that inspire further mathematical advances. Leveraging these advances, the state-of-the-art search tool \textsc{Unidom}\footnote{\url{https://github.com/billbird/unidom}}~\cite{bird2017} determined $\gamma(Q_n)$ for all $n\leq25$ and enumerated all non-isomorphic minimum queen domination solutions for boards with $n\leq18$. A more detailed discussion of prior contributions is provided in Section~\ref{sec:related-work}.

Despite the value of computational methods, they come with inherent risks regarding correctness. Since results depend on software (typically custom-written code), the potential for bugs can undermine confidence in reported outcomes. This concern is not hypothetical. Discrepancies in computational results have been documented in computer-assisted proofs of other problems, such as Lam's problem of proving the nonexistence of a projective plane of order ten~\cite{bright2021}. Even though Lam's problem was solved by two independent computational implementations~\cite{Lam1989,Roy2011}, discrepancies in the intermediate results of both implementations were later discovered, including the discovery of partial planes that were asserted to not exist but actually do~\cite{Bright2022}. During the course of this work, we likewise discovered a bug in the state-of-the-art dominating set solver \textsc{Unidom} that causes valid solutions to be missed under certain configurations. Details of a failing test case are provided in Appendix~\ref{app:unidom_bug}\@. Unfortunately, it is simply a reality of software development that even the most well-tested programs have bugs~\cite{Lam1990}. This is particularly important for code that purports to find \emph{all} solutions of a mathematical problem: how can we have confidence that all solutions were indeed found?

One promising direction for addressing this challenge is Boolean satisfiability (SAT) solving. SAT solvers have been successfully applied to a variety of difficult combinatorial problems, including the Boolean Pythagorean Triples problem~\cite{heule2016}, the chromatic packing number of $\mathbb{Z}^2$~\cite{Subercaseaux2023}, a case of the Erdős Discrepancy Conjecture~\cite{konev2014}, the minimum Kochen--Specker problem~\cite{ijcai2024p210,Kirchweger2023}, the Williamson conjecture~\cite{Bright2020}, and Ramsey theory~\cite{li2025,Barnoff2026}. Rather than implementing a specialized search algorithm, researchers encode the problem as a propositional formula. Such encodings are often considerably simpler than bespoke search software, while modern SAT solvers are highly optimized, extensively tested, and capable of generating formal resolution proofs for unsatisfiable instances. These proofs can subsequently be checked independently, providing strong guarantees that no solutions have been overlooked.

In this work, we adopt a SAT-based approach to certify all previously reported queen domination results for chessboard sizes $n\leq18$ and to resolve the previously open case $n=19$ under model enumeration, i.e., by enumerating all optimal solutions up to isomorphism. Although modern SAT solvers provide proof generation, they are general-purpose tools and do not exploit the domain-specific reasoning available to specialized solvers such as \textsc{Unidom}. Consequently, a significant part of this work focuses on designing an efficient SAT encoding, including improved literal ordering, symmetry breaking, and Cube-and-Conquer techniques to make proof-producing SAT solving competitive for this problem. Our SAT methodology is presented in Section~\ref{sec:encoding}, and its performance is evaluated in Section~\ref{sec:results}.

The main contributions of this paper are:

\begin{enumerate}
    \item We develop a proof-producing SAT framework for the queen domination problem. Our approach uses a novel line-variable encoding, a novel Hilbert curve-based literal ordering, and incorporates symmetry breaking and Cube-and-Conquer parallelization~\cite{heule2011} to improve scalability.

    \item We uncover and correct a discrepancy in the literature by showing that the number of non-isomorphic minimum queen domination solutions for $n=16$ is 371, rather than the previously reported 43.

    \item We resolve the previously open case $n=19$, showing that there are exactly 11 non-isomorphic minimum queen domination solutions.
    
    \item Our results decrease the amount of trust in computational code when compared with the previous approaches used for the queen domination problem, because all our results were accomplished by a proof-producing SAT solver that generated completeness certificates that were subsequently checked by an independent proof verifier.  In other words, our results only rely on the correctness of our SAT encoding and the correctness of the proof verifier, not the correctness of the code that did the search.
    
    \item We make our source code and the obtained solutions publicly available.
\end{enumerate}


\section{Preliminaries}
\label{sec:background}

As the core of our work relies on SAT solving, we provide a brief background on propositional satisfiability. We work with propositional formulas expressed in conjunctive normal form (CNF), which is a standard representation used in SAT solving. A formula $F$ is said to be in conjunctive normal form when $F$ consists of a conjunction (AND) of clauses, where each clause is a disjunction (OR) of literals. A literal refers to either a Boolean variable $x$ (known as a positive literal) or its negation $\neg x$ (a negative literal). A truth assignment $\alpha$ assigns each variable a value of 1 (true) or 0 (false). A literal is satisfied under $\alpha$ if the assignment makes it true—specifically, a positive literal is satisfied when the variable is assigned true, and a negative literal is satisfied when the variable is assigned false. A clause is considered satisfied if at least one of its literals is satisfied by $\alpha$, and a formula is satisfied when all of its clauses are satisfied. Such an assignment is called a model of the formula. If at least one model exists, the formula is deemed satisfiable; otherwise, it is unsatisfiable.

For a satisfiable formula \(F\), one may be interested in enumerating all models that satisfy it. This can be done by repeatedly invoking a SAT solver on \(F\), and each time a model is found adding a ``blocking clause'' to $F$ that blocks the found model from being a solution. This process continues until the extended formula \(F'\) becomes unsatisfiable, indicating that all solutions have been enumerated. For a comprehensive treatment of SAT and SAT solving, refer
to the \emph{Handbook of Satisfiability}~\cite{handbook2021}.

\section{SAT-Based Methodology}
\label{sec:encoding}

We model an \(n\times n\) chessboard as its queen graph, whose vertices correspond to board squares and whose edges connect pairs of squares sharing a row, column, diagonal, or anti-diagonal. The decision problem is whether the graph admits a dominating set of size at most \(\gamma\).

\subsection{Line-Variable Encoding}

Let \(Q_i\) denote a Boolean variable indicating whether a queen is placed on square \(i\), where squares are indexed from \(1\) to \(n^2\) in row-major order. In addition, for every row, column, diagonal, and anti-diagonal \(\ell\), we introduce a Boolean variable \(L_\ell\) indicating whether line \(\ell\) is active, that is, whether it contains at least one queen. To relate line variables to queen placements, we add, for every line \(\ell\),
\[
\neg L_\ell \vee \bigvee_{i\in\ell} Q_i.
\]
This clause ensures that whenever a line is marked active, at least one queen is placed on that line. Domination is then expressed in terms of the line variables. Let \(\mathcal{L}(i)\) denote the set consisting of the row, column, diagonal, and anti-diagonal lines passing through square \(i\). Since a square is dominated whenever at least one of these four lines contains a queen, we add, for every square \(i\),
\[
\bigvee_{\ell\in\mathcal{L}(i)} L_\ell.
\]
To bound the size of the dominating set, we impose the cardinality constraint
\[
\sum_{i=1}^{n^2} Q_i \leq \gamma.
\]
We additionally introduce the following cardinality constraint on the line variables. Since each queen belongs to exactly one row, one column, one diagonal, and one anti-diagonal, it can activate at most four distinct lines. Hence, any solution satisfying the above queen bound activates at most \(4\gamma\) lines, and we impose the cardinality constraint
\[
\sum_{\ell\in\mathcal{L}} L_\ell \leq 4\gamma
\]
where $\mathcal{L}$ is the set of all lines.


\subsection{Cardinality Encoding and Literal Ordering}

The queen and line cardinality constraints introduced in the previous section cannot be expressed directly in CNF and therefore require dedicated cardinality encodings. Among the many cardinality encodings proposed in the literature, we adopt totalizer-based encodings~\cite{Bailleux2003}. A totalizer represents a cardinality constraint as a binary tree of auxiliary counting variables, where each internal node stores the partial cardinality information of the literals in its subtree. This hierarchical representation enables strong unit propagation while producing relatively compact CNF formulas.

We employ different totalizer variants for the two cardinality constraints because they serve different purposes during solving. The queen cardinality constraint is encoded using the standard totalizer, as this encoding is compatible with the Cube-and-Conquer framework described in Section~\ref{sec:cnc}. In contrast, the line cardinality constraint is not involved in cubing. For this constraint, the modulo totalizer~\cite{Ogawa2013} achieved the best performance in our experiments and is therefore adopted.

The performance of totalizer encodings depends not only on the encoding itself but also on the ordering of the input literals, since this ordering determines the structure of the underlying counting tree and consequently influences propagation and conflict analysis~\cite{reeves2025}. For the queen cardinality constraint, we propose a novel literal-ordering strategy based on a Hilbert curve traversal of the chessboard. Because the Hilbert curve preserves spatial locality, nearby board squares tend to remain close within the totalizer tree. As a result, partial cardinality information corresponds to localized board regions, leading to stronger propagation during search. 
This ordering significantly improved the performance of the SAT solver: in the $n=14$ instance the solver was able to
show there are no dominating sets with 7 queens about 31.5 times faster using the Hilbert curve ordering over the default ordering (where variables are ordered based on their declaration order).
For the line cardinality constraint, the line variables are ordered by decreasing line length before constructing the totalizer.


\subsection{Symmetry Breaking}

To reduce redundant exploration of symmetric solutions, our encoding incorporates symmetry-breaking constraints based on a lexicographic ordering encoding proposed by Warwick Harvey~\cite{frisch2006}. Apart from the identity transformation, the \(n\times n\) chessboard has seven geometric symmetries: rotations by \(90^\circ\), \(180^\circ\), and \(270^\circ\), together with reflections across the horizontal axis, vertical axis, main diagonal, and anti-diagonal. Let \(X\) denote the Boolean vector of queen variables in the original ordering. For each non-identity symmetry, we construct the corresponding transformed vector \(Y\) and enforce \(X \leq Y\).

To encode the lexicographic comparison
\([x_1,\dots,x_N]\leq[y_1,\dots,y_N]\), where \(N=n^2\), we introduce auxiliary variables \(a_0,\dots,a_N\). Intuitively, \(a_i\) indicates whether the suffix \([x_{i+1},\dots,x_N]\) is lexicographically no greater than the corresponding suffix \([y_{i+1},\dots,y_N]\). Following Harvey's construction, for every \(0\leq i<N\), we add the clauses
\[
a_{i+1}\vee y_{i+1}\vee\neg a_i,\qquad
a_{i+1}\vee\neg x_{i+1}\vee\neg a_i,\qquad
y_{i+1}\vee\neg x_{i+1}\vee\neg a_i,
\]
together with unit clauses enforcing \(a_0\) and \(a_N\) to be true. The variables in the symmetry-breaking vectors are ordered according to the Hilbert curve, and the resulting constraints are generated for all seven nontrivial symmetries. 

Our preliminary experiments consistently showed that symmetry breaking reduced solving time, although the magnitude of the improvement varied across instances. For example, when proving that no dominating set of size \(8\) exists for the \(n=15\) instance and that no dominating set of size \(7\) exists for the \(n=14\) instance, symmetry breaking achieved speedups of \(6.64\times\) and \(5.54\times\), respectively.

\subsection{Cube-and-Conquer}
\label{sec:cnc}
For larger instances, solving the complete SAT formula directly becomes increasingly difficult. We therefore employ the Cube-and-Conquer paradigm, which partitions the SAT instance into many smaller subproblems, called \emph{cubes}, that can be solved independently by a conflict-driven SAT solver. We adopt the cubing framework of Battleman et al.~\cite{battleman2025}, which is designed for totalizer-based cardinality encodings. The framework branches on auxiliary variables from the totalizer encoding of the queen cardinality constraint. These variables encode partial counting information, allowing the search space to be partitioned according to cardinality structure rather than individual queen placements. 

In our formulation, cubing is applied only to the queen cardinality constraint because our preliminary experiments indicated that restricting cubing to this constraint provided the best overall solving performance. Specifically, following Battleman et al.~\cite{battleman2025}, cube variable selection begins at the children of the root's children in the totalizer tree. Auxiliary variables are then selected from successive depths according to the heuristic of Battleman et al.\ until the desired number of cube variables is obtained, after which all assignments to the selected variables are enumerated to generate the cubes. The resulting cubes are then solved using the SAT solver. It is worth noting that cubing also reduced solving time even when using a single CPU core. For example, for the $n=15$ instance, the solver found that no dominating set of size 8 exists about 1.93 times faster with cubing than without cubing.

\subsection{Model Enumeration, Proof Generation, and Verification}

To enumerate all optimal queen placements, we use
\texttt{cadical-exhaust},\footnote{\url{https://github.com/curtisbright/cadical-exhaust}}
an extension of the {\CaDiCaL} SAT solver for exhaustive satisfiability
solving. Rather than stopping after the first satisfying assignment,
\texttt{cadical-exhaust} enumerates all models by iteratively adding
blocking clauses until the formula becomes unsatisfiable. Upon
termination, the reported UNSAT result certifies that no additional
solutions exist.

During exhaustive search, the solver simultaneously produces a DRAT
proof in which exhaustive blocking clauses are represented as trusted
clauses~\cite{bright2020ovals}. Since the standard DRAT format does not support trusted clauses,
the resulting proofs cannot be verified by conventional DRAT proof checkers. We
therefore use \texttt{drat-trim-t},\footnote{\url{https://github.com/curtisbright/drat-trim-t}}
an extension of \texttt{drat-trim} supporting trusted clauses, to independently
verify the generated proofs. This provides machine-checkable
certificates for both unsatisfiable instances and the completeness of
the model enumeration.

\section{Results}\label{sec:results}
This section presents the main results of our study. For each instance (excluding the trivial cases $n=1$ and $n=2$), we perform model enumeration using both our SAT-based framework and the state-of-the-art specialized solver \textsc{Unidom}~\cite{bird2017}. \textsc{Unidom} is executed using a configuration that avoids the bug described in Appendix~\ref{app:unidom_bug}, i.e., by setting the split-depth parameter to 1. Experiments were carried out on the Iris HPC cluster at the University of Luxembourg using Slurm job arrays on Intel Xeon compute nodes equipped with either \(2\times\) Intel Xeon E5-2680 v4 (2.4~GHz) or \(2\times\) Intel Xeon Gold 6132 (2.6~GHz) processors. Table~\ref{tab:results} reports, for each board size, the number of non-isomorphic models, SAT enumeration time, proof verification time, proof size, and the runtime of \textsc{Unidom}.

\begin{table}[H]
\centering
\caption{Experimental results for model enumeration on queen domination instances up to $n=19$.  Times are total CPU times given in seconds.}
\label{tab:results}

\begin{tabular}{rrcccc}
\toprule

&
&
\multicolumn{3}{c}{SAT Approach}
&
\\

\cmidrule(lr){3-5}

$n$
& \# Models
& Enum.\ Time
& Verif.\ Time
& Proof Size
& \textsc{Unidom} Time
\\

\midrule

3  & 1      & 0       & 0        & 0 MB      & 0 \\
4  & 3      & 0       & 0        & 0 MB      & 0 \\
5  & 37     & 0       & 0        & 0.01 MB   & 0 \\
6  & 1      & 0       & 0        & 0.01 MB   & 0 \\
7  & 13     & 0       & 0        & 0.12 MB   & 0 \\
8  & 638    & 0       & 0        & 0.56 MB   & 0 \\
9  & 21     & 0       & 0        & 1.43 MB   & 0 \\
10 & 1      & 0       & 0        & 0.81 MB   & 0 \\
11 & 1      & 0       & 0        & 1.52 MB   & 0 \\

12 & 1      & 2.38      & 3.75       & 4.43 MB   & 2.83 \\
13 & 41     & 50        & 45         & 96.5 MB   & 69 \\
14 & 588    & 599       & 1,956      & 780 MB    & 1,812 \\
15 & 25,872 & 14,761    & 20,847     & 13.3 GB   & 47,903 \\
16 & 371    & 1,688     & 4,052      & 2 GB      & 28,909 \\
17 & 22     & 8,919     & 10,969     & 10.4 GB   & 20,369 \\
18 & 2      & 5,051     & 8,040      & 5.9 GB    & 21,271 \\
19 & 11     & 171,575   & 252,706    & 86.5 GB   & 714,733 \\

\bottomrule
\end{tabular}
\end{table}

Table~\ref{tab:results} shows that the proposed SAT framework consistently outperforms the specialized solver \textsc{Unidom} on all nontrivial benchmark instances. In addition, our results uncover a discrepancy in the previously reported enumeration for $n=16$. While the literature~\cite{bird2017} reports 43 non-isomorphic solutions, our proof-producing SAT approach establishes that there are in fact 371.
The reason for the discrepancy is unclear, though it may be a result of a bug that is present in \textsc{Unidom} under certain configurations (see Appendix~\ref{app:unidom_bug}). Furthermore, our work provides the first complete enumeration of the $n=19$ instance, establishing that it admits exactly 11 non-isomorphic solutions. For all remaining board sizes, our results agree with those previously reported in the literature.

These results also demonstrate that the line-based encoding we introduce in this paper significantly improves the performance of the SAT solver. 
In a preliminary version of this paper~\cite{rostami2025}, our SAT encoding did not use the line variables $L_\ell$ and the solver took about 22.8 times longer to
solve and verify the instance for $n=19$.

\section{Related Work}
\label{sec:related-work}

The queen domination problem belongs to a broader family of domination problems on chessboards that differ in board geometry, chess piece, placement restrictions, and domination variant~\cite{hedetniemi2020}. Among these, the minimum queen domination problem on the standard square chessboard has received considerable attention. Research has focused not only on determining the minimum number of queens required to dominate the board, but also on enumerating all optimal solutions, often considered up to board symmetries or isomorphism~\cite{bird2017}.

Previous work on the minimum queen domination problem can be broadly divided into theoretical and computational research. On the theoretical side, researchers have established increasingly stronger upper and lower bounds on the queen domination number, either for arbitrary board sizes or for specific values of $n$~\cite{weakley1995,weakley2002,burger1997}. Together, these results often narrow the possible value of $\gamma(Q_n)$ to within one queen. Reasoning about queen domination in terms of rows, columns, diagonals, and anti-diagonals has also played an important role in theoretical investigations. Classical lower-bound arguments exploit structural properties of these lines~\cite{cockayne1990}, while more recent work has introduced line-based relaxations of the queen domination problem to simplify analysis and strengthen lower bounds~\cite{karandikar2023}. 

In parallel with these primarily theoretical investigations, researchers have also developed specialized computational methods to determine exact values of $\gamma(Q_n)$ and enumerate all optimal solutions. Gibbons and Webb~\cite{gibbons1997} incorporated isomorphism rejection into a backtracking search to solve several previously open instances. Kearse and Gibbons~\cite{kearse2000} subsequently investigated three complementary search strategies based on queen placements, undominated squares, and attacking sets, selecting the most effective strategy for different regions of the search space according to empirical performance. More recently, Bird developed the highly optimized \textsc{Unidom} solver~\cite{bird2017}, which combines specialized branching heuristics, bounding techniques, efficient data structures, distributed computation, and several implementation optimizations to resolve numerous open instances, including complete model enumeration for board sizes $n\leq18$.

\section{Conclusion}
\label{sec:conclusion}

In this paper, we complete the model enumeration of the queen domination problem up to isomorphism for all $n \leq 19$. We achieve this through a logic-based approach that encodes the problem into propositional logic. Given the natural formulation of the queen domination problem as a SAT instance, this approach avoids the need for implementing a specialized search algorithm from scratch. Furthermore, it generates correctness certificates proving that the model enumeration is complete, i.e., that no models up to isomorphism have been missed. These certificates can be verified independently using third-party proof checkers. In our experiments, all generated proofs were verified using \texttt{drat-trim-t}~\cite{bright2020ovals}. Consequently, the correctness of our results depends only on the correctness of the SAT encoding and the proof checker, rather than on the correctness of the SAT solver itself.

We further improve our approach by introducing auxiliary variables for lines and a novel literal-ordering strategy based on the Hilbert curve, which preserves spatial locality and substantially outperforms existing ordering strategies. Combined with symmetry breaking and a Cube-and-Conquer workflow, these techniques enable our SAT approach to outperform the leading specialized solver for the queen domination problem, \textsc{Unidom}.

Finally, our work corrects errors in the existing literature on the queen domination problem. During the course of this work, we discovered a bug in the solver \textsc{Unidom}, which had previously been used to exhaustively enumerate all solutions for $n \leq 18$. We also show that the $n=16$ instance has 371 non-isomorphic optimal solutions rather than the previously reported 43. Furthermore, we resolve the previously open $n=19$ instance by showing that it has exactly 11 non-isomorphic optimal solutions.

More broadly, this work demonstrates that careful SAT encoding design can substantially improve proof-producing combinatorial search. In particular, the ideas developed in this work---introducing auxiliary variables that explicitly capture problem structure, strengthening the formulation with global constraints over those variables, and exploiting geometry-aware literal ordering---may also prove useful for other graph domination problems and, more generally, for proof-producing SAT applications.

\section*{Code and Data Availability}
The implementation is available at \url{https://github.com/TahaRostami/Gamma/tree/gamma-plus}. The data is
available at \url{https://zenodo.org/records/21193115}.

\section*{Acknowledgement}

We would like to thank Bill Bird for generously sharing the \textsc{Unidom} source code and for many helpful discussions regarding its implementation. We also thank Wendy Myrvold for helpful discussions on implementation techniques and distributed search, and Ali Gholami for valuable comments on an earlier version of this manuscript.

\bibliographystyle{unsrt}
\bibliography{references}

\appendix
\section{Bug in \textsc{Unidom}}
\label{app:unidom_bug}

\textsc{Unidom} is a fast, specialized solver for domination problems that has recently been used to resolve several previously open cases, including the largest board sizes reported in the literature to date~\cite{bird2017}. While \textsc{Unidom} is a highly sophisticated and well-optimized tool, it lacks support for proof generation. As a result, using it without caution introduces the risk that undetected bugs may compromise the correctness of its output.

In this section, we present a test case that demonstrates such a bug in \textsc{Unidom} under certain settings. This example illustrates how subtle implementation flaws may lead to incorrect results and underscores the importance of certificate generation in computational investigations of combinatorial problems.

\textsc{Unidom} supports distributed computation based on a divide-and-conquer paradigm, which is frequently used to handle larger instances (e.g., $n \geq 15$). In this mode, nodes at the upper levels of the search tree (up to a given cutoff depth) are duplicated across all processes, meaning that each process independently explores this part of the tree. Beyond this depth, the remaining subtrees are partitioned and assigned to specific processes. This distributed mode is parameterized by three user-specified values: the number of jobs, the job ID, and the cutoff depth. While the first two depend on the available computational resources, the cutoff depth must be manually selected.

Our tests revealed a bug in \textsc{Unidom}'s distributed mode. We instructed the tool to compute all solutions for a specific board size, then aggregated the solutions into a \emph{frequency matrix}, where each entry indicates how often a queen appears at that square across all solutions, normalized by the total number of solutions found. Given the inherent symmetry of the chessboard, the resulting matrix should itself be symmetric. Any deviation from this symmetry indicates a flaw in the computation. We reported this issue to the author of \textsc{Unidom}, who agreed that our analysis was sound.


Our attempts to fix the bug were unsuccessful, but experiments revealed that it is specifically tied to \textsc{Unidom}’s strongest bounding strategy when used with larger cutoff depths. When the cutoff depth was set to small values such as 0, 1, or 2—even with the same bounding strategy—\textsc{Unidom} passed our symmetry test. For example, we tested combinations with 1, 5, and 10 processes and cutoff depths of 0 and 1, and in all cases obtained consistent, symmetric results. 

\begin{figure}[ht]
    \centering
    \includegraphics[width=0.40\linewidth]{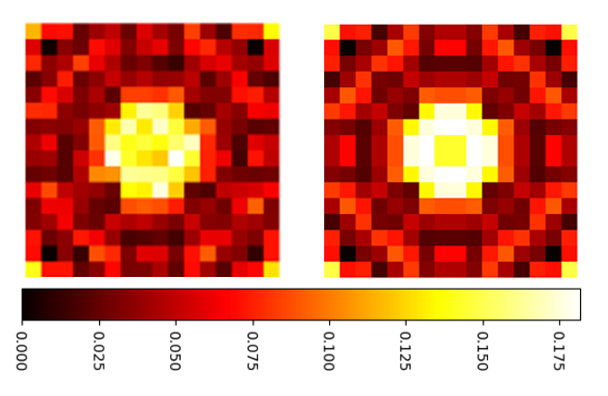}
    \caption{
Heatmaps of a $16 \times 16$ board: \textbf{Left} shows incorrect behavior due to a bug in \textsc{Unidom}; \textbf{Right} shows correct output with a small cutoff depth.
    }
    \label{fig:symmetry-failure-and-pass}
\end{figure}

Figure~\ref{fig:symmetry-failure-and-pass} illustrates this contrast for $n=16$: the left heatmap displays the asymmetric frequency matrix resulting from the bug when using a large cutoff depth, while the right heatmap shows the symmetric frequency matrix obtained with a small cutoff depth where \textsc{Unidom} operates correctly.

\end{document}